\documentclass[epj,final]{svjour}
\usepackage{graphicx}
\usepackage{epsfig}
\usepackage{color}
\usepackage{amsmath,amssymb}
\usepackage{midfloat}
\usepackage{here}

\begin{document}

\authorrunning{P. Kl\"upfel et al}
\title{Systematics of collective correlation energies from
  self-consistent mean-field calculations}
\author{P. Kl\"upfel\inst{1}  \and J. Erler\inst{1}  \and P.--G. Reinhard\inst{1}
\and J. A. Maruhn\inst{2}
} 
\institute{ 
Institut f{\"u}r Theoretische Physik, Universit{\"a}t Erlangen,
Staudtstrasse 7, D-91058 Erlangen, Germany
\and
Institut f{\"u}r Theoretische Physik, Universit{\"a}t Frankfurt,
Max-von-Laue-Str. 1, D-60438 Frankfurt a.~M., Germany 
}
\date{\today / Received: date / Revised version: date}
\abstract{ 
The collective ground-state correlations stemming from low-lying
quadrupole excitations are computed microscopically.  To that end, the
self-consistent mean-field model is employed on the basis of the
Skyrme-Hartre-Fock (SHF) functional augmented by BCS pairing. The
microscopic-macroscopic mapping is achieved by quadrupole-constrained
mean-field  calculations which are processed further in the
generator-coordinate method (GCM) at the level of the Gaussian overlap
approximation (GOA).
We study the correlation effects on energy, charge radii, and surface
thickness for a great variety of semi-magic nuclei. A key issue is to
work out the influence of variations of the SHF functional. We find
that collective ground-state correlations (GSC) are robust under
change of nuclear bulk properties (e.g., effective mass, symmetry
energy) or of spin-orbit coupling. Some dependence on the pairing
strength is observed. This, however, does not change the general
conclusion that collective GSC obey a general pattern and that their
magnitudes are rather independent of the actual SHF parameters.
} 

\maketitle

\section{Introduction}

The nucleus is a highly correlated many-body system and thus the topic
of correlations has always accompanied nuclear physics.  Short-range
correlations dominate due to the huge short-range repulsion in the
nucleon-nucleon interaction. Their treatment requires very involved
approa\-ches as, e.g., Brueckner-Hartree-Fock, variational Jastrow,
hyper-netted chain, or no-core shell model calculations, see
e.g. \cite{Dic92aR,Rei94aR,Pan97aR,Nav00a,Fuc06a}.  Large-scale
applications employ simpler approaches, among which the most
microscopic ones are self-consistent mean-field models based on
effective energy-density functionals or effective forces,
respectively. The functional is motivated by a formal derivation from
an underlying many-body approach, see e.g. \cite{Neg72a}, but adjusted
phenomenologically because of the as yet insufficient input from
ab-initio models; for a recent review see \cite{Ben03aR}.  A widely
used functional is, e.g., the Skyrme-Hartree-Fock (SHF) approach.
However obtained, the nuclear energy-density functional is supposed to
contain all crucial correlations effectively. The reasoning is quite
similar to that of density-functional theory in electronic systems
where theorems and clean theoretical developments have come much
farther, for an overview see \cite{Dre90aB}.  The perfect functional
thus allows describing a system by a mere mean-field calculation and
yet obtaining the correct energy and density.  That ideal functional
is known to be highly non-analytic to account for the sudden changes
from one particle number to another \cite{Dre90aB}. Actual
functionals, however, are rather smooth functions of the involved
densities, partially owing to approximations in their derivation and
partially for practical reasons of manageability.  This, in turn,
means that the functional can incorporate only the smooth trends of
the correlations and will probably fail to account for quickly
changing contributions. Yet such effects are known to exist in the
nuclear landscape.  The nuclear shell structure leads to dramatic
changes in the nuclear shape when adding or removing nucleons, known
as the nuclear Jahn-Teller effect \cite{Rei84a,Kos95b}. Pairing
moderates these strong fluctuations \cite{Rei84a,Wer94a}, but sizeable
fluctuations remain as can be seen from the changes in the low-energy
spectra along isotopic or isotonic chains \cite{Ram87aER}.  The lowest
excitation in even-even nuclei is in most cases a quadru\-pole state
assigned as $2_1^+$. This $2_1^+$ state is a collective state. It is
associated with substantial recoupling of the simple mean-field
excitations which, in turn, produces significant correlation
effects. These collective correlations are not smooth with changing
particle number and thus are not included in the energy-density
functional. They have to be added explicitly. That has been discussed
for decades, but mostly based on semi-empirical estimates, see
e.g. \cite{Rei79a,Bar85b}.  Thorough calculations based on
self-consistent mean-field models have also been tried for a long
time; see e.g. \cite{Gir82a,Gir82b,Bon90b,Bon91a}. {Extensive
surveys} running over all nuclear landscapes based on SHF have been
published recently {\cite{Ben05a,Ben06a,Sab07a} and there
exists also a systematic study using the Gogny force \cite{Ber07e}
showing very similar trends.}  It is the aim of this paper to give a
survey of collective correlations within the SHF mean field, exploring
in detail the dependence on the SHF functional. The aim is to prepare
ground for a revision of the phenomenological adjustment of SHF by
finding out the least correlated nuclei.

The low-lying quadrupole states can be formulated in terms of the
Bohr-Hamiltonian, which establishes collective dynamics in the five
quadrupole degrees of freedom \cite{Boh52}. The parameters of the
collective Hamiltonian are usually adjusted phenomenologically, see
e.g. the applications in \cite{Gne71} and more recently in the
interacting boson model \cite{Iac87aB}.
The direct connection between a microscopic description and a collective
picture can be established by a collective path, {which
consists of a continuous series of mean-field states with prescribed
deformations, produced by a quadrupole constraint.} The mapping of the
path into collective dynamics is established by virtue of the
generator coordinate method (GCM).  \cite{Hil53a,Gri57a}.  The complex
integral equations of the GCM can be simplified using the
Gaussian-Overlap-Approximation (GOA) which, furthermore, allows to
establish contact between the microscopic foundation and the
collective Bohr-Hamiltonian \cite{Rei87aR,Bon90b}.  In practice, there
are, on the one hand, fully fledged GCM calculations which skip the
collective Hamiltonian as intermediate level and compute low-energy
spectra directly from the coherent superposition of the collective
path; these sophisticated calculations imply exact projection for the
conserved quantities, like particle number, angular momentum, and
center of mass; from the many published results we mention here
\cite{Val00a,Rod02a} as two recent examples. On the other hand, there
are the techniques which use the Bohr-Hamiltonian as an intermediate
stage, for an early example see \cite{Gir82a,Gir82b} and for more
recent achievements \cite{Liber99,Pro04a}.  We are going here to use
the GCM-GOA. The path is generated from axially deformed SHF-BCS
states. The full five-dimensional topology is regenerated by
interpolation into the triaxial plane in an extension of the method
presented in \cite{Fle04a}. The interpolation scheme promises reliable
results for vibrators whose collective deformation rarely reaches large
values. It is thus well suited for the present survey dealing with chains
of semi-magic nuclei.

\section{Formal framework}
\label{sec:model}

\subsection{The Skyrme mean-field model}
The starting point for the self-consistent microscopic description
is the SHF energy functional 
$E=E(\rho,\tau,{\cal J},{\bf j},{\bf\sigma})$,
which is expressed in terms of a few local densities and currents
obtained as sums over single-particle wave functions: density $\rho$,
kinetic density $\tau$, spin-orbit density ${\cal J}$, current ${\bf
j}$, and spin density ${\bf\sigma}$, each occurring twice, once
for protons and once for neutrons.  It is augmented by a pairing
functional deduced from a zero-range two-body force (DI=delta
interaction) or combined with a density dependence
(DDDI=density-dependent delta interaction) \cite{Ben00c}. 
{
They are described with the pairing functional
\begin{equation}\label{eqn:pairfunct}
E_{\rm pair} = -\sum_{q\in p,n} \frac{V_{0q}}{4}\int\!d^3 {\bf r} 
\left[1 - \frac{\rho({\bf r})}{\rho_0}\right]\Re\{\chi_q({\bf r})\}^2
\end{equation}
where DDDI-pairing corresponds to $\rho_0=0.159\,$fm$^{-3}$, while 
DI-pairing is recovered for $\rho_0\to\infty$.}
{The pair density is 
$\chi({\bf r})=\sum_\alpha u_\alpha v_\alpha|\varphi_\alpha({\bf r})|^2$.
The real part thereof confines the contributions to time-even
parts. This coincides with the standard pairing functional 
for stationary states (where $\chi$ is purely real) and it
implies that time-odd pairing effects in dynamical response
are ignored.}
%
Pairing is treated within the BCS approximation and
augmented by the Lipkin-Nogami (LN) correction which
serves to force a minimal amount of pairing everywhere and so prevents
the sudden changes associated with the pairing phase transition.
That smoothing is compulsory for the further processing of
the mean-field states in the description of collective dynamics.
{ The BCS approximation is applicable for well bound nuclei
where the Fermi energies are safely below the continuum threshold.
This holds for the nuclei considered here along the whole
collective deformation path.}
For details about SHF, BCS and LN, see the review
article \cite{Ben03aR}.
%

The mean-field equations are derived variationally from the given
energy functional. They determine the ground state or a locally stable
isomer. Collective motion goes through a succession of deformed mean
fields, the collective path.  The low-lying quadrupole mode is
particularly soft and the collective path is related to quadrupole
deformed shapes.  We generate the path by imprinting a dedicated
deformation using quadrupole-constrained mean-field equations,
often called constrained Hartree-Fock (CHF),
\begin{subequations}
\begin{eqnarray}
  \hat{H}_{\rm MF}|\Phi_{\alpha_{20}}\rangle 
  &=&
  {\cal E}|\Phi_{\alpha_{20}}\rangle
  \quad,
\label{eq:Cmfeq}
\\
  \hat{H}_{\rm MF}
  &=&
  \hat{h}_{\rm MF}-\epsilon_{\rm F}\hat{N}-\lambda\hat{Q}_{20}
  \quad,
\label{eq:CmfHam}
\end{eqnarray}
where $\hat{h}_{\rm MF}$ is the SHF-BCS mean field Hamiltonian
(depending on the local densities and thus on the state
$|\Phi_{\alpha_{20}}\rangle$), $\epsilon_{\rm F}$ is the Fermi energy,
and $\lambda$ the Lagrange parameter for the quadrupole constraint.
$\hat{Q}_{20}$ is the quadrupole operator and $\alpha_{2m}$ its
dimensionless expectation value, i.e.
\begin{eqnarray}
  \hat{Q}_{2m}
  &=&
  r^2Y_{2m}f_{\rm cut}({\bf r})
  \quad,
\\
  \alpha_{2m}
  &=&
  \frac{4\pi}{5}
  \frac{\langle\Phi_{\alpha_{20}}|r^2Y_{2m}|\Phi_{\alpha_{20}}\rangle}
       {Ar^2}
  \quad,
\label{eq:label}
\end{eqnarray}
\end{subequations}
with $A$ the total particle number and $r$ the r.m.s. radius.  The
index $m$ can run over $-2,-1,0,1$ and $2$. For a while, we will
consider only $m=0$ which corresponds to axially symmetric
deformations.  The states are labeled with the dimensionless
quadrupole moment (\ref{eq:label}). The quadrupole operator is
modified by a {Wood-Saxon-like} damping function 
$f_{\rm cut}$ {with an extension of three times the
nuclear radius and width of 1 fm in order to suppress the
unbound regions from the asymptotics $\propto -x^2, -y^2$, or
$-z^2$}  \cite{Rut95a}.  The equations
are solved with an extra iterative loop to maintain a wanted value of
$\alpha_{20}$ \cite{Cus85a}. This is done for a dense set of
deformations $\alpha_{20}$. At the end we have the collective path
$\{|\Phi_{\alpha_{20}}\rangle\}$ as a series of mean-field states
along which the collective motion can evolve. Its further processing
will be discussed in section \ref{sec:collham}.
%
Before carrying on, we ought to mention that the optimal scheme for
generating a collective path is adiabatic time-dependent Hartree-Fock
(ATDHF) which generates the constraints in a self-consistent manner
without imposing a preconceived idea of the collective deformation,
see e.g. \cite{Rei87aR,Goe78a}. Experience shows that the quadrupole
constraint is a good approximation in cases where the low lying
$2_1^+$ state is a truly collective mode. That applies practically to
all nuclei, except for the doubly-magic ones. Thus our results near
doubly-magic nuclei have to be taken with care.

The SHF functional sets only a framework within which there is a
manifold of different parameterizations.  Most available parameter
sets describe ground states properties equally well but differ in
other observables like, e.g., excitation spectra or nuclear matter
properties \cite{Ben03aR}. One thus should consider several
sufficiently different parameterizations to distinguish
particular features of a given parameterization from general features of
SHF.
We will consider the following standard Skyrme parameterizations:
SkM$^*$ as a widely used traditional standard \cite{Bar82a}, Sly6 as a
recent fit which includes information on isotopic trends and neutron
matter \cite{Cha97a}, and SkI3 as a recent fit which maps the
relativistic iso-vector structure of the spin-orbit force
\cite{Rei95a}.  The set contains distinct effective masses where
SkM$^*$ has $m^*/m=0.8$ while SLy6 and SkI3 have significantly lower
values (0.69, or 0.6 respectively).  SkI3 differs in that it has
basically no proton-neutron coupling in the spin-orbit mean
field. Thus all three forces differ somewhat in the actual shell
structures they produce.  Besides the effective mass, the bulk
parameters (equilibrium energy and density, incompressibility,
symmetry energy) are comparable.
In addition, we employ a set with a systematic variation of effective
mass, symmetry energy and spin-orbit force \cite{Rei99a} to
explore the sensitivity to these aspects of the force.  

\subsection{The collective Hamiltonian}
\label{sec:collham}

Large-amplitude collective motion proceeds along the collective path,
a series of mean-field states which differ very little in energy.  Its
description requires a coherent superposition of the states along the
collective path. That is done in the framework of the GCM. The GOA
allows to map the microscopic picture into a collective Hamiltonian
for the five quadrupole degrees of freedom
$\boldsymbol\alpha=
(\alpha_{2-2},\alpha_{2-1},\alpha_{20},\alpha_{21},\alpha_{22})
$
\cite{Rei87aR}. The final result is summarized briefly.
%

The collective Schr\"odinger equation to be solved reads
\begin{equation}
  \left(
    \hat{H}^{\rm coll}
    -
    \delta\epsilon_\mathrm{F}
    \hat{N}^{\rm coll}
  \right)
  \phi(\boldsymbol\alpha)
  =
  E_n^{\rm coll}\phi(\boldsymbol\alpha)
  \quad,
\label{eq:collSE}
\end{equation}
with the collective Hamiltonian 
\begin{equation}\label{eqn:COLLHAM}
  \hat{H}^{\rm coll} 
  = 
  -\frac{1}{4}\left\{\nabla_\mu, \left\{
  B^{\mu\nu}({\boldsymbol\alpha}), 
  \nabla_\nu\right\}\right\} + V({\boldsymbol\alpha})
  \quad.
\end{equation}
It is obtained from the general rules of a collective map
which apply analogously to the collective map $\hat{N}^{\rm
coll}$ of the particle-number operator \cite{Rei87aR}. The details are given
here for the Hamiltonian.  
%
The potential in $\hat{H}^{\rm coll}$
is given by the mean-field energy from which the spurious
contributions from rotational, vibrational, and center-of-mass
zero-point energies are subtracted, i.e.
\begin{subequations}
\begin{eqnarray}
  V({\boldsymbol\alpha}) 
  &=& 
  E_{\rm SHF}({\boldsymbol\alpha})  -  
  E_{\rm ZPE}({\boldsymbol\alpha})  -
  E_{\rm cm}({\boldsymbol\alpha})
  \quad,
\label{eq:collpot}\\
  E_{\rm ZPE}({\boldsymbol\alpha})
  &=&
  \frac{1}{2} \lambda_{\mu\nu} B^{\mu\nu}
  - 
  \frac{1}{4} (\lambda^{-1})^{\mu\nu}\nabla_\mu\nabla_\nu
              E_{\rm SHF}
  \quad,
\\
  \lambda_{\mu\nu}({\boldsymbol\alpha}) 
  &=&
  \langle \Phi_{{\boldsymbol\alpha}} | 
    \left\{\hat{{\cal P}}_\mu,\hat{{\cal P}}_\nu\right\}
  | \Phi_{{\boldsymbol\alpha}} \rangle 
  \quad,
\label{eq:collmass}\\
  {\hat{\cal P}}_\mu 
  |\Phi_{{\boldsymbol\alpha}} \rangle 
  &=&
  \mathrm{i} \frac{\partial}{\partial\alpha^\mu} 
  |\Phi_{{\boldsymbol\alpha}} \rangle \ .
\label{eq:collmom}
\end{eqnarray}
\end{subequations}
where the ${\hat{\cal P}}_\mu$ are the generators of a change in 
the quadru\-pole momentum.
The inverse collective mass tensor is evaluated 
as
in terms of the
constraint mean-field Hamiltonian $\hat{H}_{\rm MF}$ and the 
\begin{subequations}
\begin{equation}
  B^{\mu\nu}({\boldsymbol\alpha}) 
  =
  \frac{1}{2}\langle \Phi_{{\boldsymbol\alpha}} | 
    \left[\hat{{\cal Q}}^\mu,\left[\hat{H}_{\rm MF},\hat{{\cal Q}}^\nu\right] 
-  \mathrm{i}
    \hat{h}_{\rm resp,{\cal Q}}
\right]
  | \Phi_{{\boldsymbol\alpha}} \rangle 
\label{eq:collinvmass}
\end{equation}
where $\hat{\cal Q}^\mu$ are the generators of a dynamical boost.
They are determined self-consistently as dynamical linear response to
the collective momentum $\hat{\cal P}_\nu$, i.e.
\begin{equation}
  \left(
    \mathrm{i}\left[\hat{H}_\mathrm{MF},{\hat{\cal Q}}^\mu\right] + 
    \hat{h}_{\rm resp,{\cal Q}}\right)
  |\Phi_{{\boldsymbol\alpha}} \rangle 
  =
  2B^{\mu\nu}{\hat{\cal P}}_\nu
  |\Phi_{{\boldsymbol\alpha}} \rangle 
  \quad,
\label{eq:crankATDHF}
\end{equation}
\end{subequations}
where $\hat{H}_\mathrm{MF}$ is the mean-field Hamiltonian belonging to
state $|\Phi_{{\boldsymbol\alpha}} \rangle $ and 
$\hat{h}_{\rm resp,{\cal Q}}$ is the response Hamiltonian for
time-odd perturbations about  $|\Phi_{{\boldsymbol\alpha}} \rangle $
\cite{Rei92b}.
The mass thus obtained is called self-consistent or ATDHF mass. 
It is to be noted that most straightforward GCM calculations omit the
dynamical linear response and deal with an approximate mass which can
be deduced purely from the static configurations. That amounts to
replacing the dynamical boost operator  $\hat{\cal Q}^\mu$ in
formula (\ref{eq:collinvmass}) simply by the redundant complement
to  $\hat{\cal P}^\mu$ \cite{Rei87aR}, i.e. 
\begin{equation}
  \hat{\cal Q}^\mu|\Phi_{{\boldsymbol\alpha}} \rangle
  = 
  -\mathrm{i}(\lambda^{-1})^{\mu\nu}{\hat{\cal P}}_\nu
  |\Phi_{{\boldsymbol\alpha}} \rangle
  \quad.
\label{eq:crankGCM}
\end{equation}
{ Accounting for the dynamical response explores the dynamic
path for slow collective motion. Within the assumptions of the GOA
framework it thus corresponds to a {dynamic
generator-coordinate method} (DGCM).}  We will compare later on these
two approaches to the collective mass.
The collective path $|\Phi_{{\boldsymbol\alpha}} \rangle$ is computed
for axially symmetric shapes, i.e. along $\alpha_{20}$. This
determines the raw collective potential $E_{\rm SHF}$, the masses
$B^{00}$ and the widths $\lambda_{00}$ along the $\alpha_{20}$. By a
slight rotation of the nucleus, generated by the angular momentum
operator $\hat{J}_x$ or $\hat{J}_y$, the moments of inertia are
evaluated providing additional information about rotational properties
of the nucleus.  Including this information allows to evaluate the
$B^{1,-1}$ and $\lambda_{1,-1}$ components of the GOA tensors.  The
data from prolate ($\alpha_{20}>0$) and oblate ($\alpha_{20}<0$)
configurations are interpolated into the triaxial plane with methods
similar to those employed in \cite{Fle04a} but now accounting for the
topology of the five dimensional quadrupole configuration space (see
appendix \ref{sec:appx}).  The requirement of rotational symmetry for
the collective Hamiltonian $\hat{H}^{\rm coll}$ and of smoothness in
the triaxial direction allows to determine all needed ingredients of
the collective Hamiltonian in the full triaxial plane, for the regime
of small deformations.  It is to be noted that the assumption of
small deformations limits the present treatment to nearly spherical
nuclei.  These are the ones which we will study in the following.

Coming back to the collective Schr\"odinger equation
(\ref{eq:collSE}), we see that it contains a constraint on particle
number. Each mean-field state $|\Phi_{\alpha_{20}}\rangle$ is tuned to
the same average particle number $\langle\hat{N}\rangle=N$ by 
tuning the Fermi energy in the mean-field Hamiltonian
(\ref{eq:CmfHam}). But the coherent superposition may lead to a
small drift of the average particle number. The term
$\propto\delta\epsilon_\mathrm{F}$ in (\ref{eq:collSE}) serves
for a final particle-number correction (PNC). 

\section{Results and discussion}

\subsection{Ground-state correlations of semi-magic nuclei}

\begin{figure*}
\centerline{\includegraphics[width=15.6cm]{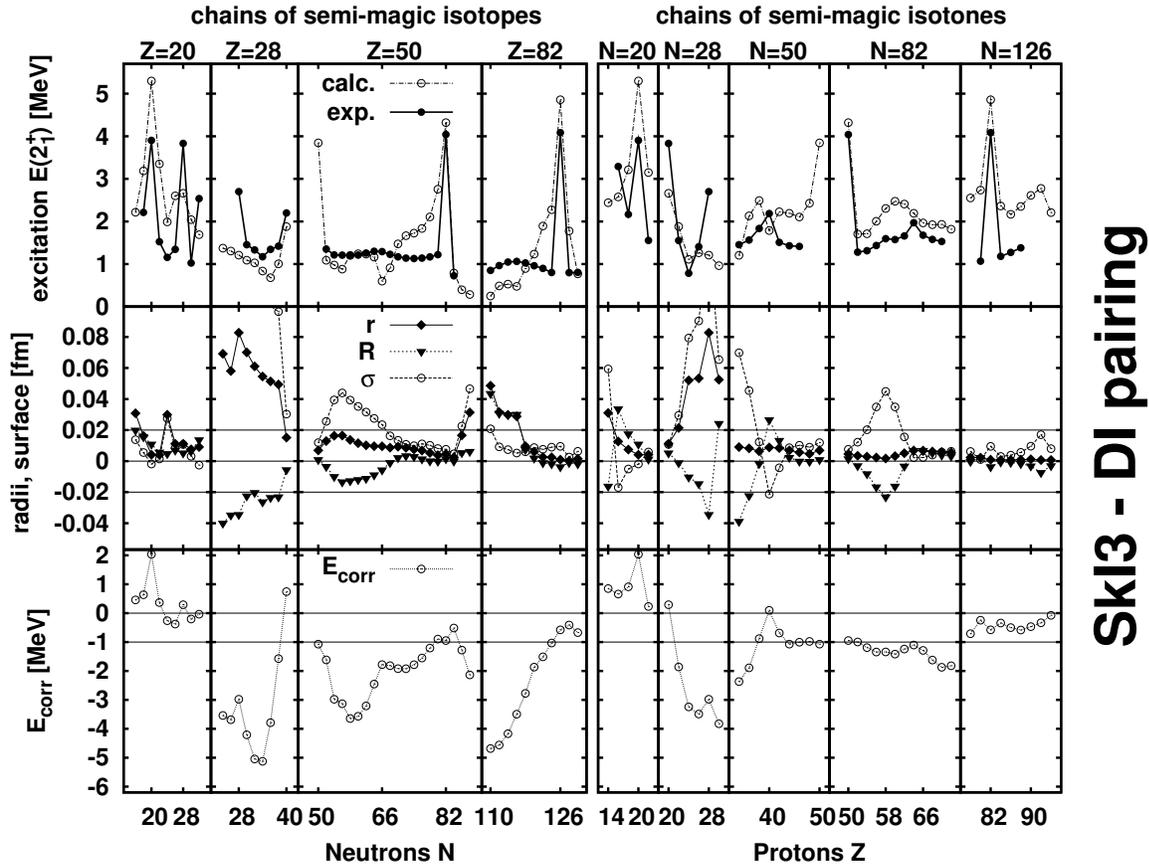}}
\caption{\label{fig:corr_e2_all_ski3_DI_PNC_bw}
  Correlation effects on semi-magic isotopic (left) and isotonic
  (right) chains computed with SkI3 and DI pairing.
  Upper panels: $E_2^+$ excitation energies, theoretical and
  experimental as indicated.
  Middle panels: correlation shifts of charge r.m.s. radii $r$, 
  diffraction radii $R$,
  and surface thicknesses $\sigma$.
  Lower panels: correlation energies.
}
\end{figure*}
Figure~\ref{fig:corr_e2_all_ski3_DI_PNC_bw} shows the results of the
SHF plus DGCM-GOA calculations for the correlation effects on the four
bulk observables energy $E$, charge r.m.s. radius $r$, diffraction
radius $R$, and surface thickness $\sigma$ for all semi-magic nuclei
considered in this survey. The upper panel complements that by the
energies of the low lying $2^+$ compared with the experimental
results {(taken from \cite{Nudat})}.  
That figure is generic in the sense that the trends seen
here are typical for a great variety of Skyrme forces. It summarizes
the correlation effects from low lying $2^+$ states.

The uppermost panels show the $2_1^+$ excitation energies. The heavier
systems distinguish nicely the doubly magic nuclei by large
$E(2^+_1)$. The experimental energies drop suddenly when going away
from the doubly magic stage while the theoretical results show a
somewhat softer transition. That qualitative mismatch is cause by two
approximations which are not optimal for doubly-magic nuclei: the LN
recipe for stabilized pairing and a simple quadrupole constraint
rather than full ATDHF. {It is to be noted that the same
effect was found in the study of \cite{Sab07a} which differs in
several details of the treatment but does also employ a collective
path from CHF.} There are slight quantitative discrepancies in the
mid-shell regions. These change with the Skyrme force and can be
related to the level density at the Fermi surface. Such variations
will be discussed later.

The correlation effects (middle and lower panel) show the expected
trends, being minimal at doubly magic points and growing large at
mid-shell. The question of interest is how large they actually become.
The horizontal lines indicate typical values of acceptable
uncertainties in a well fitted Skyrme force, 1 MeV for the energy and
about 0.02 fm for charge radii or surface thickness.
The lighter systems generally have larger correlation effects, often
beyond the desired bounds. Some Z=20 and N=20 systems even
have slightly positive correlation energies, with $^{40}$Ca being the
worst case.  That unphysical effect is due to the GOA. It requires
smooth collective paths \cite{Rei76a}. But for these small semi-magic
systems, the abrupt pairing phase transition lies well inside the
range of the shape fluctuations of the ground state and it is not
sufficiently well smoothed by the LN scheme. An
$E_\mathrm{corr}\approx +0.5\,\mathrm{MeV}$ is anyway within the
expected precision of the method. The $^{40}$Ca is more dramatic
because the low $2^+$ state is not really collective.
Besides these critical light nuclei,
correlation effects are negative for energies and positive for
r.m.s. radii, as it should be. The parameters derived from the charge
form factor, diffraction radius and surface thickness can go both ways
and do so. The results carry two surprises:
first, the form parameters (middle panels) generally show smaller
correlation effects relative to the goal than the energies,
and second, the correlation energies are large within isotopic chains
while remaining moderately small in isotonic chains. 
Both features call for a revision of fitting strategies.
A thorough discussion of reliable data sets for the adjustment of Skyrme
parameters will follow in a separate publication. Here we continue
with discussing variations of Skyrme forces, pairing, and collective
approximations to corroborate the above result. To that end, we
confine the presentation to the isotopic Z=50 chain and the isotonic
N=82 chain. That contains all necessary trends while rendering the
figures more transparent.
 
\subsection{Low lying $2^+$-excitations}

\begin{figure}
\centerline{\includegraphics[width=8.8cm]{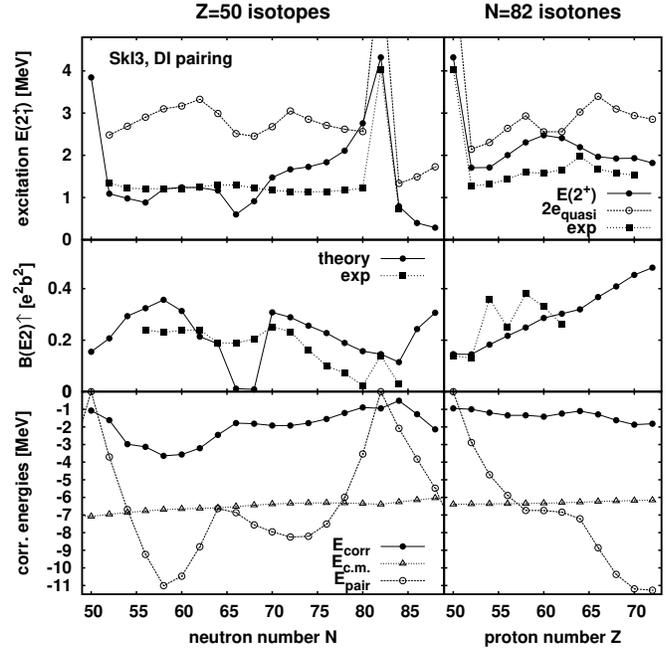}}
\caption{\label{fig:equasi_ecorr_BE2_ski3_bw}
  Collectivity of the low lying $2^+$ states for Sn isotopes (left)
  and $N=82$ isotones (right) 
  computed with SkI3 and DI pairing.
  Upper panels: comparison of $E(2^+_1)$ excitation energies, from pure mean
  field (two quasiparticle energies), full collective calculations, and 
  experiment.
  Middle panels: $B(E2)\!\!\uparrow$ values from collective calculations 
  and experiment.
  Lower panels: correlation energies.
}
\end{figure}
Figure \ref{fig:equasi_ecorr_BE2_ski3_bw} shows more details about the
$2^+_1$ state and the correlations energies.  The upper panel 
contains the $E(2_1^+)$ excitation energies and adds the
two-quasiparticle ($2QP$) energies $E_{2QP}$ for comparison.  The
difference between the $E_{2QP}$ and the fully dressed $E(2_1^+)$
characterizes the strength of the residual interaction which is
clearly related to the correlation (lowest panels).
One also sees the systematic effect that isotones
(upper right) have less collective downshift than isotopes (upper
left), and thus smaller correlation energy (lowest panels).

The middle panel complements the excitation energies by the
corresponding B(E2) values. {These agree in the average fairly
well with the experimental data (taken from
\cite{PhysRevC.47.392,PhysRevC.61.024309,vaman:162501,bart03,gabl01,radf04,radf05,speid93,bazz91}).
However, there are substantial deviations in details: a dramatic drop
at N=66 and 68 and a trend to overestimation towards shell closure
N=82.}  { The breakdown of the $B(E2)$ in the mid-shell tin
isotopes is related to the} kink in the $E(2_1^+)$ which, in turn, can
be traced back to a kink in the two-quasiparticle energy (upper
panel).  The reason is the interference with a shape transition which
is probably unrealistic, an artefact of that particular force (we see
it also for SLy6, but not for SkM$^*$ and the other forces discussed
here).  {The differences around the doubly-magic tin isotope
result from using the CHF path as an approximation to the optimized
ATDHF path. By construction CHF overestimates the collectivity of the
ground state and thus also the transition probability to the
$2^+$-state. The mismatch is {similarly seen in the too large}
excitation energies. The same artefact {appears also in}
studies based on different nucleon-nucleon interactions using a
similar collective model based on a CHF path \cite{Sab07a} {
while a recent QRPA study performs much better in that respect
\cite{Ter06a}. As QRPA can be considered as the small amplitude limit of
ATDHF, that result indicates} that full ATDHF will cure the mismatch
for those isotopes.}

The lowest panels show the c.m. correlation energies and pairing
energies in addition to the collective correlation energy. The
c.m. energy has a very smooth trend because it is related to
excitation across shells. The collective correlation energies, on
the other hand, are related to intra-shell excitations and thus
changing from isotope to isotope. The trend is very similar to
the trend of the pairing energy which depends similarly on the
available intra-shell phase space.

\begin{figure}
\centerline{\includegraphics[width=8.8cm]{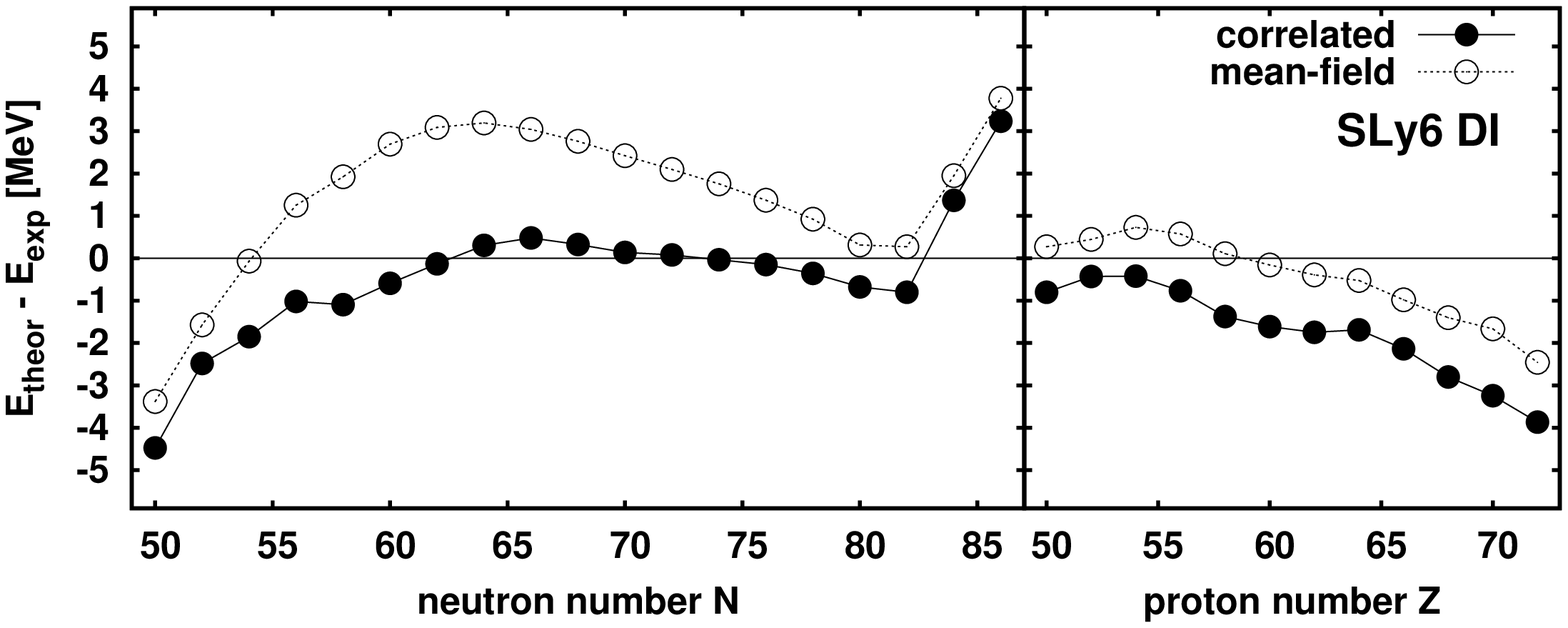}}
\caption{\label{fig:compare_exp_DGCM_bw}
  Difference between theoretically computed and experimental
  binding energies in Z=50 isotopes (left) and N=82 isotones (right)
  computed with SLy6 and DI pairing.
  Compared are pure spherical mean-field calculations (open circles)
  with correlated ground state energies (filled circles).
}
\end{figure}
%
Figure \ref{fig:compare_exp_DGCM_bw} demonstrates the effect of the
collective ground state correlations (GSC) on the trend of the 
deviation from experimental binding
energies. The pure mean-field calculations (open circles) display an
unresolved trend for the isotopic chain (left) while having already
the correct trend for the isotones (right). The GSC now perfectly
compensate that due to strong correction for the mid-shell isotopes
(see figure \ref{fig:equasi_ecorr_BE2_ski3_bw}). The practically
flat trend of the GSC for the isotones leaves their already agreeable
pattern intact.

\subsection{Comparing approximations}

\begin{figure}
\centerline{\includegraphics[width=8.8cm]{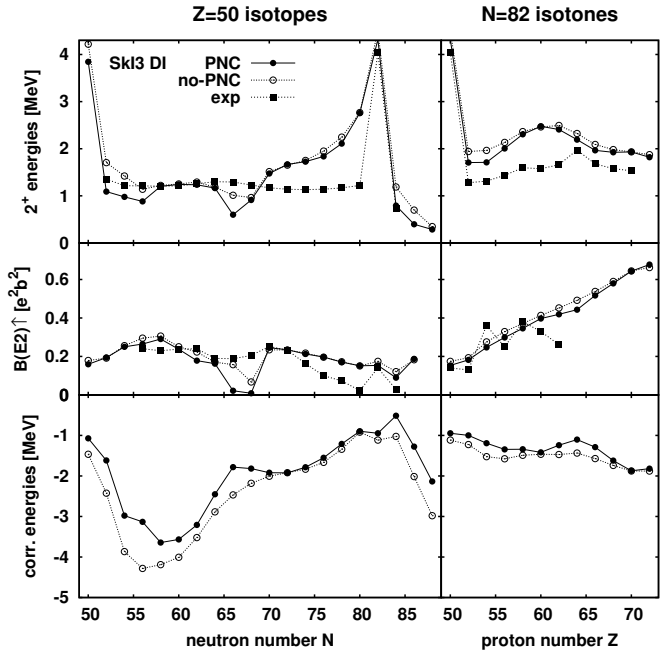}}
\caption{\label{fig:Energies_ski3_DI_NOPNC_bw}
  Effect of the particle-number correction on excitation  energies
  (top) and on correlation  energies (bottom) for
  Sn isotopes (left)
  and $N=82$ isotones (right) computed with SkI3 and DI pairing.
}
\end{figure}
Figure \ref{fig:Energies_ski3_DI_NOPNC_bw} shows the effect of
particle number correction (PNC) on excitation and correlation energies. 
%
The case without PNC means to ignore the term
$\propto\delta\epsilon_\mathrm{F}$ in (\ref{eq:collSE}).
%
The effect on $E(2_1^+)$
is small, but the PNC are slightly improving the steepness 
of the transition from doubly-magic nuclei to semi-magic ones.
The effect on correlation energies is still small but noteworthy at a
quantitative level. It is particularly interesting that it always
reduces $E_\mathrm{corr}$ while the general trends remain unchanged.
{The PNC reacts particularly sensitive at N=66 and 68,
the region where shape fluctuations lead to the much
reduced B(E2) values.}

\begin{figure}
\centerline{\includegraphics[width=8.8cm]{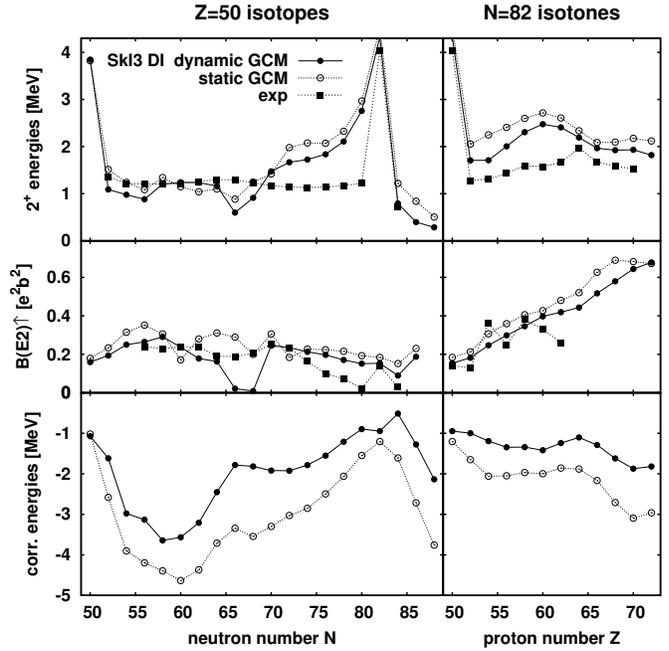}}
\caption{\label{fig:Energies_ski3_DI_cranking_bw}
  Effect of the dynamical response (``dynamic GCM'') on excitation  energies
  (top) and on correlation  energies (bottom) for 
  Sn isotopes (left)
  and $N=82$ isotones (right) computed with SkI3 and DI pairing.
}
\end{figure}
Figure \ref{fig:Energies_ski3_DI_cranking_bw} deals with the treatment
of the collective mass, comparing results of the (standard) GCM which
employs only stationary deformation paths yielding the redundant mass
(\ref{eq:crankGCM}) with the dynamic GCM which explores the dynamical
response of the nucleus and so employs the self-consistent cranking
mass (\ref{eq:crankATDHF}). The dynamic GCM lowers the $E(2^+_1)$
and improves, in particular, the steepness for the step away from
doubly-magic nuclei.  
{The B(E2) values
react particularly sensitive at N=66 and 68,
precisely the region which is very sensitive for some forces.
}

For the correlation energies, the trends remain
robust qualitatively, but there is a sizeable reduction by about
20-50\%. 
{Both trends can be explained by the feature that the 
dynamical response is a first step towards the fully self-consistent
ATDHF path (that is why self-consistent cranking is often called
ATDHF cranking). The variation in the larger space reduces the
excitation energies and, at the same time, takes a different cut
through the collective landscape which eventually reduces
correlations \cite{Rei79b}.
}
This shows the importance of proper dynamical response
{and indicates that full ATDHF will improve the
  performance at and around doubly magic nuclei.}

\subsection{Variation of the Skyrme- and pairing force}

\begin{figure}
\centerline{\includegraphics[width=8.8cm]{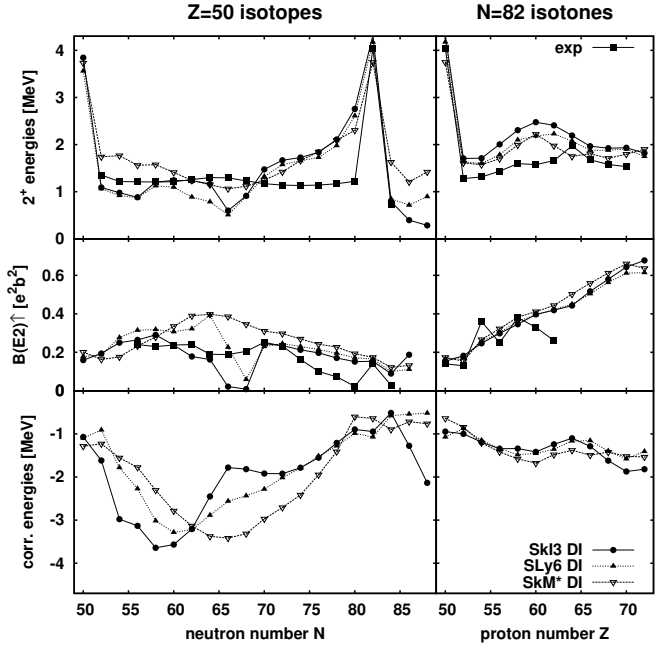}}
\caption{\label{fig:Energies_all_forces_DI_bw}
  Correlation energies and $2^+$-energies calculated with SkI3, SLy6 and
  SkM$^*$ and DI-pairing.
}
\end{figure}
Figure \ref{fig:Energies_all_forces_DI_bw} compares results for three
widely used Skyrme parameterizations. The global trends of the results
are the same: smaller $E_\mathrm{corr}$ for isotonic chains, overall
size of $E_\mathrm{corr}$, growth of $E_\mathrm{corr}$ towards mid
shell, and too smooth decrease away from doubly-magic nuclei. There
are differences in detail. The dip in the $E(2_1^+)$ 
{and B(E2)} at $^{116}$Sn
does not appear for SkM$^*$, and this force also has the correlation
maximum at a different position in the isotopic chain. 

It is not possible to judge which of the features of the three forces
is most responsible for the slight differences. In order to
disentangle the driving agents, we now are considering a series of
Skyrme parameterizations produced under the same conditions
while varying systematically one selected feature. It originated from a
systematic analysis of giant resonances and their dependence on the
Skyrme parameters \cite{Rei99a}.
\begin{figure}
\centerline{\includegraphics[width=8.8cm]{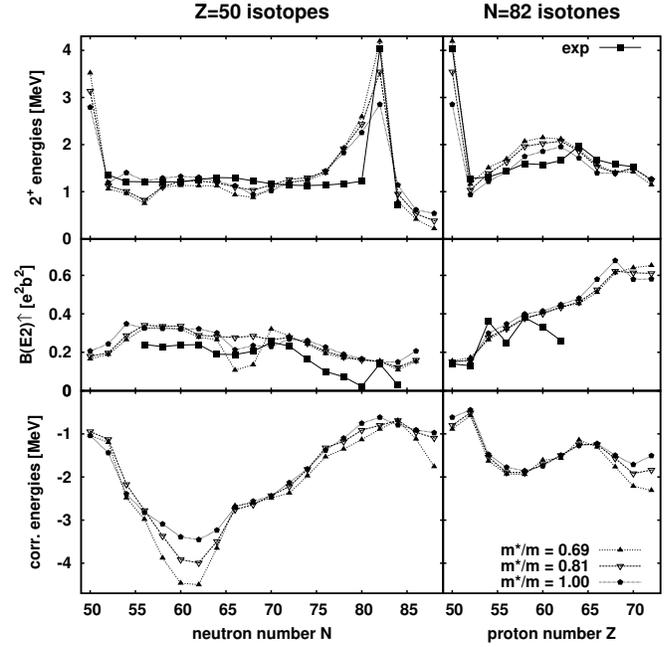}}
\caption{\label{fig:Energies_noref_mass_bw}
  Influence of the effective nucleon mass $m^*/m$  on excitation  energies
  (top) and on correlation  energies (bottom) for
  Sn isotopes (left)   and $N=82$ isotones (right).
  Compared are results for three Skyrme forces with systematically 
  varied $m^*/m$ while keeping all other features the same.
}
\end{figure}
Figure \ref{fig:Energies_noref_mass_bw} shows a variation of the
effective mass $m^*/m$. This feature has an impact on the shell
structure, mainly the level density. The $E(2_1^+)$ are almost 
inert.
{The B(E2) values are also generally inert, but show
some sensitivity around the notoriously critical N=66.
The same holds for
$E_\mathrm{corr}$ where
some effects
can be seen in regions
around N=60 and N=82 for  Sn isotopes and here the correlations
increase with decreasing $m^*/m$ as one would expect. The other
regions where $E_\mathrm{corr}$ is less sensitive are dominated by
pairing as we will see later.
}
\begin{figure}
\centerline{\includegraphics[width=8.8cm]{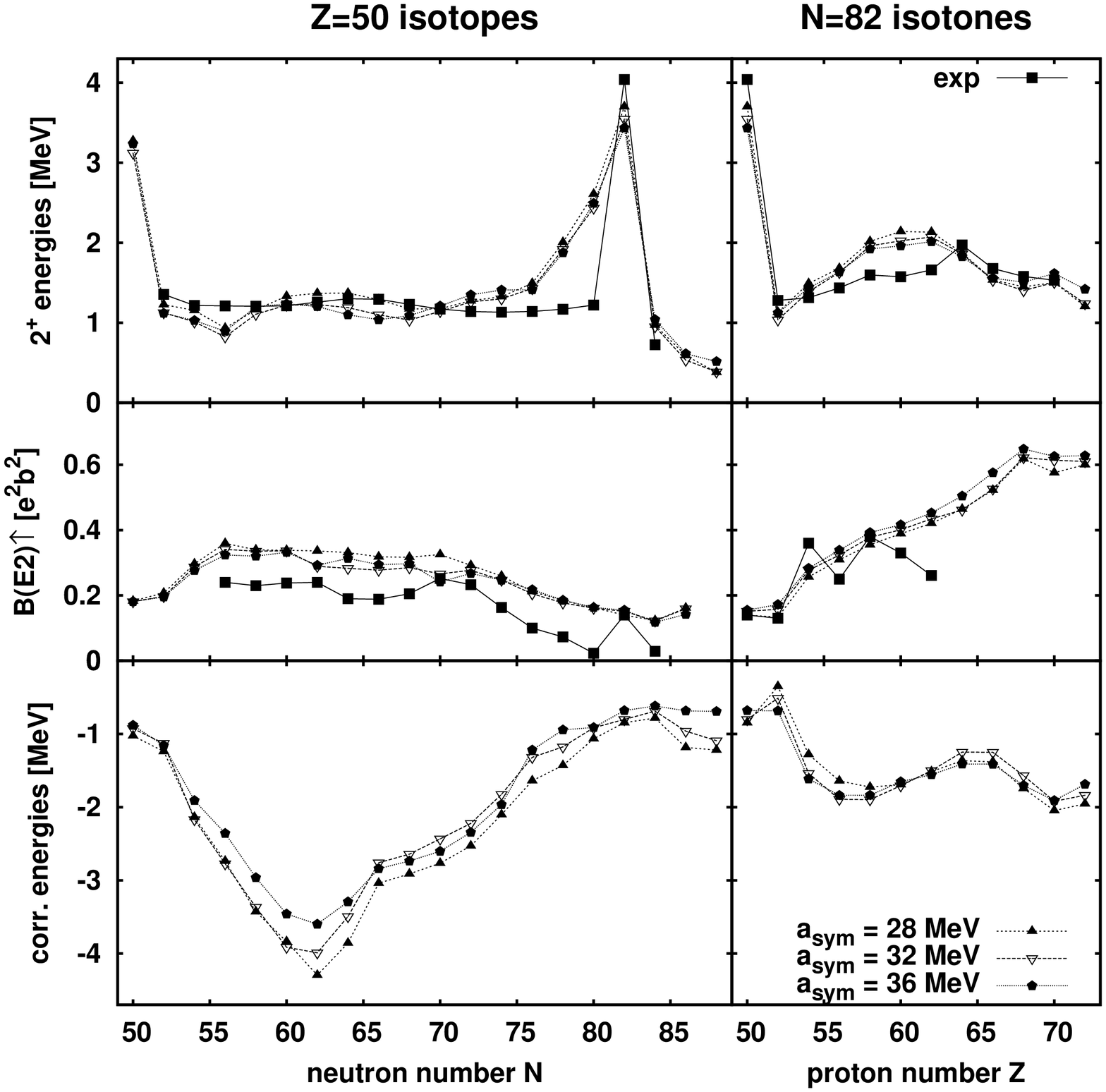}}
\caption{\label{fig:Energies_noref_asym_bw}
  Influence of the symmetry energy $a_\mathrm{sym}$ on excitation energies
  (top) and on correlation  energies (bottom) for
  Sn isotopes (left)   and $N=82$ isotones (right).
  Compared are results for four Skyrme forces with systematically 
  varied $a_\mathrm{sym}$ while keeping all other features the same.
}
\end{figure}
The variation of the symmetry energy shown in figure
\ref{fig:Energies_noref_asym_bw} is {very} robust and shows
{even} smaller effects.

We also studied other variations of bulk parameters (sum rule
enhancement {$\kappa$}, incompressibility, density dependence of
$a_\mathrm{sym}$). Changes in those cases were even smaller.

\begin{figure}
\centerline{\includegraphics[width=8.8cm]{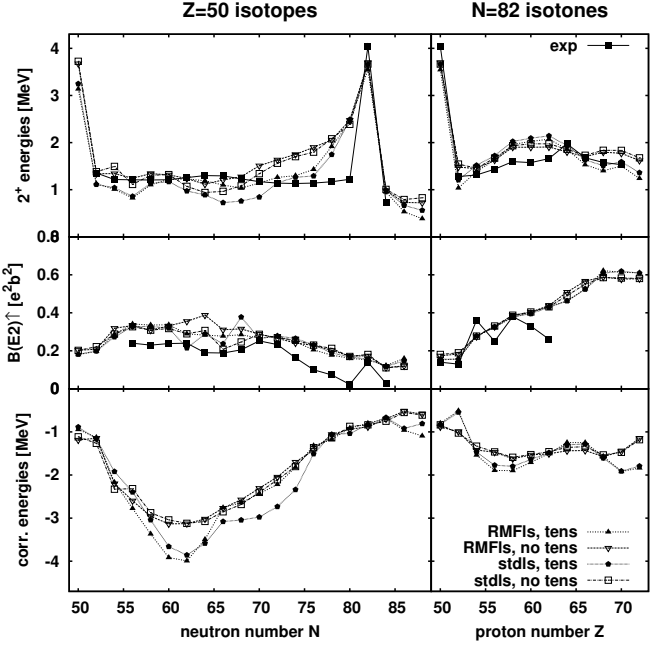}}
\caption{\label{fig:Energies_noref_ls_bw}
  Influence of the form of the l$*$s functional  on excitation  energies
  (top) and on correlation  energies (bottom) for
  Sn isotopes (left)   and $N=82$ isotones (right).
}
\end{figure}
Figure \ref{fig:Energies_noref_ls_bw} shows the effect of using
different forms for the l$\cdot$s coupling
{\cite{Rei99a}}. Recall that there is some
variety in handling the l$\cdot$s forces for SHF functionals.  The standard
form is derived from a zero-range two-body l$\cdot$s interaction
\cite{Vau70a} which leads to a fixed mix of isoscalar and isovector
l$\cdot$s coupling, the isovector being half as large as the isoscalar one.
That is indicated by ``stdls'' in the figure. The classical limit of
the relativistic mean-field model suggests that the isovector
spin-orbit should be nearly zero \cite{Rei95a}, a variant which is
indicated by ``rmfls'' in the plot. Moreover, there is a spin-orbit
term $\propto {\cal J}^2$ emerging from the kinetic zero-range
interaction. It is called tensor l$\cdot$s term. Many parameterizations omit
that term. Inclusion is indicated in the figure by ``tens'' and
omission by ``no tens''. 
$a_\mathrm{sym}$ 
{The impact of the
different l$\cdot$s models is well visible: Inclusion of tensor l$\cdot$s
yields somewhat larger correlation energies and lower $E(2^+)$ at
places, particularly in combination with the standard l$\cdot$s model.
The  B(E2), on  the other hand, are robust towards shell closures
and show some sensitivity mid shell.
It is obvious  that the  l$\cdot$s force has a large effect on shell
structure. We observe that changing the  l$\cdot$s model can even
change the ordering of levels near the Fermi energy. This, in turn,
modifies quasi-particle energies and thus  $E(2^+)$ as well as
$E_\mathrm{corr}$.
}

\begin{figure}
\centerline{\includegraphics[width=8.8cm]{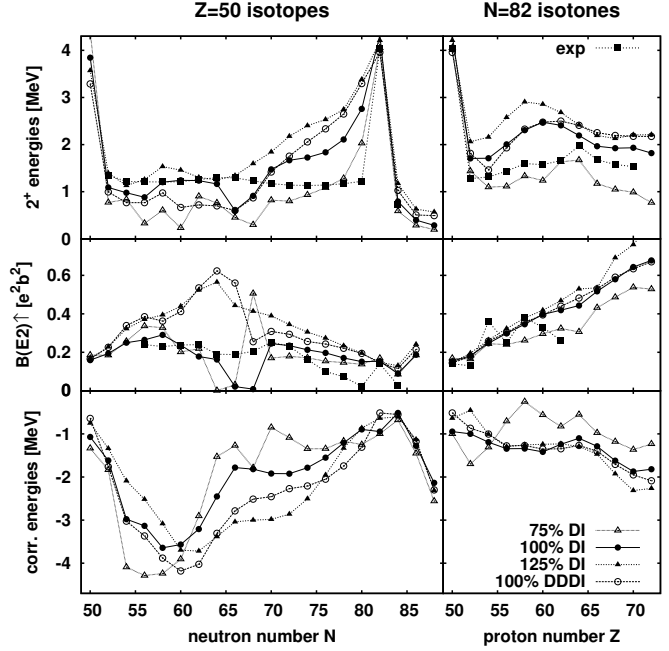}}
\caption{\label{fig:Energies_ski3_DI_pairing_bw}
  Influence of the pairing strength and pairing functional on excitation  energies
  (top) and on correlation  energies (bottom) for
  Sn isotopes (left)   and $N=82$ isotones (right).
  Compared are results for DI pairing with systematically varied
  strength and DDDI pairing, all for SkI3.
}
\end{figure}
Figure \ref{fig:Energies_ski3_DI_pairing_bw} demonstrates the effect
of pairing. 
{
It shows a variation of the strengths $V_{0q}$ of the 
pairing functional (\ref{eqn:pairfunct}) for the case of DI-pairing
relative to the original DI and DDDI pairing functional. 
}
The effects are considerable {for all three observables
shown}. Again, the $E(2_1^+)$ are very sensitive due to a close
breakdown of pairing for the lower strength.  It is noteworthy that
DDDI even produces different trends than DI for $E(2_1^+)$. The
correlation energies are more robust, yet showing here the largest
variations of all cases studied here.  The results for the $E(2_1^+)$
in comparison to experiment show that the reduced pairing strength is
clearly ruled out. What remains proves again that the trends of
correlation energies and their approximate magnitudes are a very
robust quantity, a generic feature of all reasonable SHF functionals.

\subsection{Comparison with a large-scale GCM analysis}

\begin{figure}
\centerline{\includegraphics[width=8.8cm]{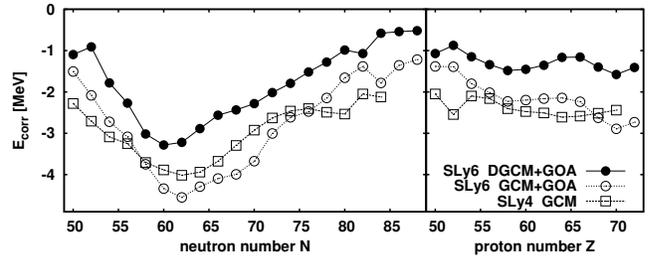}}
\caption{\label{fig:compare_GCM_bw}
  Comparison of different approximations to the
  quadrupole correlated state. The DGCM-GOA and GCM-GOA calculations
  made use of the topological interpolation to recover the full quadrupole
  configuration space (5D) during the solution of the GCM-GOA Hamiltonian.
  The GCM calculations are performed directly by coherent superposition of
  the microscopic states along the static axial path (3D) avoiding the
  intermediate step of the GOA approximation and the collective 
  Schr\"odinger equation.
}
\end{figure}
Figure \ref{fig:compare_GCM_bw} tries a comparison with the GCM
calculations of \cite{Ben05a,Ben06a}. It has to be taken with care
because there are several differences at once: The method of
\cite{Ben05a,Ben06a} handles pairing in a slightly different phase
space (cutting symmetrically above and below the Fermi energy), it
deals with full particle number projection (where we restore particle
number in the average), {it uses full angular momentum
projection (while we do that in GOA),}
it uses only the stationary path,
{it performs the collective superposition only within the
space of axially symmetric shapes (where we interpolate to fully
triaxial collective dynamics),} and it circumvents the collective
Schr\"odinger equation by determining the superposition weights
directly through the Hill-Wheeler equation with the given overlap
kernels. Finally, we deal with two slightly different forces, SLy4 and
SLy6.  But these forces belong to the same family and behave very
similar such that this should not spoil the comparison.  In spite of
all differences, the results in figure \ref{fig:compare_GCM_bw} are
quite instructive. First of all, there is an overall agreement in
trends and magnitude. Looking more closely, one sees even a close
agreement between the full, but static, GCM treatment of
\cite{Ben05a,Ben06a} and our approach at the same level, namely the
one using the mere GCM mass (denoted ``GCM'' in the plot).  The main
difference seems to come from the self-consistent mass in DGCM.
{That statement, however, has to be taken with care because 
the multitude of differences in treatment leaves many options open.
It would require a very detailed study to disentangle the various
ingredients. This is beyond the scope of the present paper.}

\subsection{A rough estimate of the average correlation energy}

\begin{figure}
\centerline{\includegraphics[width=8.8cm]{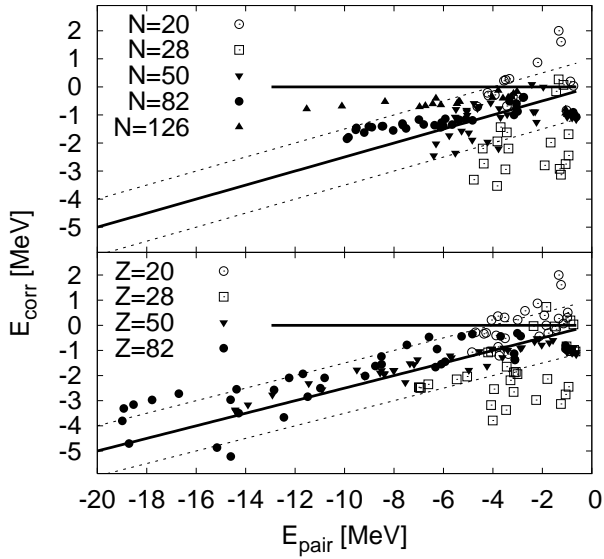}}
\caption{\label{fig:Ecorr_Epair_semi}
Correlation diagram between $E_\mathrm{corr}$ and the pairing
energy $E_\mathrm{pair}$ of the static mean-field ground-state 
for isotonic chains (upper panel)
 and isotopic chains (lower panel). The correlation and pairing
energies were calculated using the Skyrme parameterizations
SkI3, SLy6 and SkM$^*$ and DI-pairing.  
The average linear trend $E_{\rm corr} = 0.25 E_{\rm pair}$ 
is indicated as the solid line through the scatter while 
typical deviations of $\pm1$MeV are indicated by the dashed lines.
 }
\end{figure}
The question arises whether one can establish a relation between the
correlation energy and other quantities which are simpler to
compute. Pairing properties are the most promising candidates for that
purpose because pairing is switched on and off with similar trends as
collective correlations. We have figured out that the closest
connection exists with the pairing energy {(\ref{eqn:pairfunct})
of the mean-field ground state}. 
Figure \ref{fig:Ecorr_Epair_semi} collects results from a variety of forces
in a diagram $E_\mathrm{corr}$ versus $E_\mathrm{pair}$. The upper
panel shows isotopic chains (i.e. neutron pairing) 
and the lower isotonic ones (i.e. proton pairing ) as indicated.
The nuclei with large pairing and correlation energies indicate a
clear linear trend which we have determined as
$E_\mathrm{corr}=0.25\,E_\mathrm{pair}$ by least squares fitting.
Small nuclei (open symbols) show a larger scatter, but have smaller
correlation energies throughout. The Ni isotopes with Z=28 and the
isotones with N=28 are particularly off the line. 
The fit line is accompanied by two parallel lines indicating the
band $\pm 1$ MeV. It is gratifying to see that most nuclei with
$N,Z\geq 50$ stay fairly well within that band.  That nourishes the
hope that one may develop a simple formula for previewing correlation
effects. It ought to be mentioned, however, that the trend is not so
strictly linear for DDDI pairing. This, as well as the strong
deviations for the N=126 isotones here, have yet to be understood.

\begin{figure}
\centerline{\includegraphics[width=8.8cm]{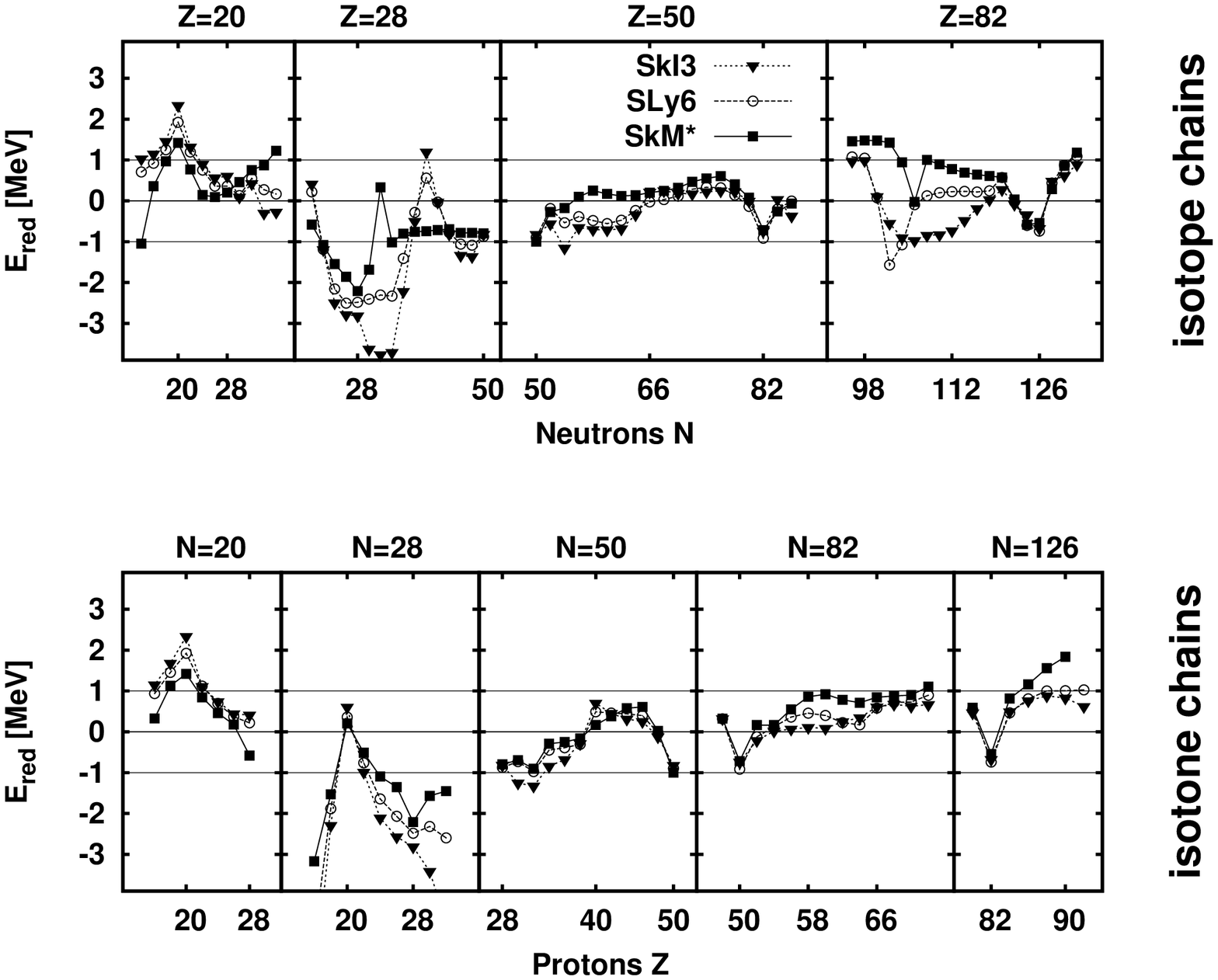}}
\caption{\label{fig:Ered_semi}
The reduced correlation 
$E_{\rm red}=E_{\rm corr} - 0.25 E_{\rm pair}$
 drawn versus the proton 
number $Z$ for the semi-magic isotone chains
(upper panel) or neutron number $N$ for the semi-magic isotones 
(lower panel). As in figure \ref{fig:Ecorr_Epair_semi} the data
were obtained with the three Skyrme parameterizations 
SkI3, SLy6 and SkM$^*$ and DI pairing.
  }
\end{figure}
Taking the simple trend from figure \ref{fig:Ecorr_Epair_semi}, we
define a reduced correlation energy as 
$E_\mathrm{red}=E_\mathrm{corr}-0.25\,E_\mathrm{pair}$.
Figure \ref{fig:Ered_semi} shows the trends of $E_\mathrm{red}$ over
all isotopic and isotonic chains. That shows even better the nice
performance for the heavier nuclei and some yet unresolved trends.
The latter {call yet for refinement of the estimate 
and, at the same time, may serve to give a clue for the improvement.}

\section{Conclusions and outlook}

Based on nuclear mean-field calculations using the Skyrme
energy-density functional, we have studied ground state correlations
as produced by the soft modes associated with low-lying quadrupole
states. The collective dynamics was described by a coherent
superposition of deformed mean-field states which were produced by
quadrupole constrained Skyrme-Hartree-Fock calculations. 
The information on
collective transport (masses, inertia) was evaluated by a
self-consistent calculation of the dynamical response to changing
deformation. The thus correlated state was handled with the
generator-coordinate method (GCM) in the Gaussian overlap
approximation (GOA) yielding effectively a collective Hamiltonian in
quadrupole coordinates and similar mapping of other observables. The
study concentrated on semi-magic nuclei with more or less shape
fluctuations about the spherical shape. The microscopic mean-field
calculations employed axially symmetric deformations while the
collective dynamics was considered with all five quadrupole degrees of
freedom by properly interpolating the collective properties
(potential, masses and quantum corrections) into the triaxial plane.
The procedure was employed for a systematic exploration of ground
state correlations on nuclear bulk properties (energy, radii, surface
thickness) scanning its dependence on the underlying energy
functional and on detailed corrections as, e.g., restoration of
particle number or dynamical response. We have also investigated the
energies and B(E2) values of the low-lying $2^+$ excitation.

On the side of the model development, we find that the proper
treatment of dynamical response has moderate effect on $E(2^+_1)$
excitation energies (about 10--20\%) but a huge one on correlations
(factor two on $E_\mathrm{corr}$). The restoration of the correct
particle number makes less effects, a few percent on $E(2^+_1)$ and
about 10\% on $E_\mathrm{corr}$. We also find that the low lying
quadrupole state is a truly large-amplitude mode. Even in doubly magic
$^{208}$Pb, a small amplitude limit turns out to be unreliable (15\%
effect on $E(2^+_1)$).
Some small nuclei cause problems in that they have a positive
correlation energy. The GOA becomes insufficient there due to a small
number of nucleons involved in the motion (lack of collectivity) and due
to a strong impact of the pairing phase transition.
For system sizes, the trends of $E(2^+_1)$ do not reproduce the
experimentally found steep decrease when going away from doubly-magic
nuclei. This defect was also found in earlier calculations using GCM.
That is a problem of the simple quadrupole constraint used for
generating the collective path. More elaborate schemes as, e.g.,
adiabatic time-dependent Hartree-Fock are required to describe
correctly the transition from collectivity in mid-shell nuclei 
to predominantly two-quasiparticle excitations in doubly-magic systems.

The large scale variation of energy functionals has shown that the
correlation effects are very robust. Their trends and magnitudes come
out quite similarly for all Skyrme forces in this survey.  Even the more
sensitive $2^+$ excitation shows very similar patterns throughout.
Differences are seen, of course, in detail. Variation of bulk
properties as symmetry energy or effective mass has little effect.
Somewhat more sensitivity is seen for variations of the spin-orbit
model, in particular tensor spin-orbit can produce up to 20\% on
$E_\mathrm{corr}$ and 40\% on $E(2^+_1)$. The strongest effect comes
from variation of the pairing strength. Particularly the $E(2^+_1)$ are
very sensitive which is no surprise because of known strong interplay
between shape transitions and pairing. However, the basic result of
the robustness of $E_\mathrm{corr}$ remains valid even here. 
We have also found a close relation between $E_\mathrm{corr}$ and the
pairing energy {of the mean-field ground state} 
which show in the average a linear dependence on each
other.

The results provide clear indications for the next steps.  On the
formal side, one needs to go for the self-consistent constraints
according to ATDHF replacing the
simple-minded quadrupole constraint and one should search for a better
smoothing of the pairing phase transition, if possible.  For further
applications, one can take advantage of the robust information on
correlation effects and use these as guide for searching least
correlated ground state observables as input for better adjustments of
Skyrme energy functionals. It is also worth while to improve the estimate
of correlations energies in terms of simple to compute quantities.
Work in all these directions is in progress.

\section*{Acknowledgments}
That work was supported by the BMBF contract number 06 ER 124D.  We
thank M.~Bender, T.~B\"urvenich and J.~Erler for helpful
discussions. We thank G. Hager
and the Regional Computing Center Erlangen for help in performing the
large computing tasks.

\begin{appendix}

\section{\label{sec:appx}Interpolating the Collective Hamiltonian}
\subsection{Intrinsic system}
The Cartesian representation of the quadrupole coordinates ($\alpha_\mu$) is 
advantageous for formulating the GCM-GOA Hamiltonian (\ref{eqn:COLLHAM}). 
For the solution
of the collective Schr\"o\-dinger equation a formulation in terms of the intrinsic 
coordinates $(\beta,\gamma,{\boldsymbol\vartheta})$ is more convenient 
\cite{Boh52,Kum67,Gne71}:
\begin{equation}
\alpha^\mu
          = \beta\left\{ \cos(\gamma) D^{2*}_{\mu,0}({\boldsymbol\vartheta}) + 
                          \sin(\gamma) 
                          {\textstyle{
  \frac{D^{2*}_{\mu,+2}({\boldsymbol\vartheta}) + 
        D^{2*}_{\mu,-2}({\boldsymbol\vartheta})}{\sqrt{2}}}} \right\} \ .
\end{equation}
The shape of the nucleus is parameterized in one of the principal 
axes systems of the nucleus
by the total deformation $\beta$ and triaxiality $\gamma$ while the orientation
of this principal axis system is given by the three Euler angles 
$\boldsymbol\vartheta$.
The representation of the collective Hamiltonian in these coordinates is advantageous 
due to the decoupling of rotation and vibration modes. In detail the collective 
Hamiltonian reads
\begin{subequations}
\begin{equation}
\hat{H} = \boldsymbol\nabla^\dagger {\bf B}(\beta,\gamma) \boldsymbol\nabla + V(\beta,\gamma) \ ,
\end{equation} 
with collective potential $V$, mass tensor
\begin{equation}
{\bf B}=\left(\begin{array}{ccccc}
B_{\beta\beta}  & B_{\beta\gamma} & 0 & 0 & 0 \\
B_{\beta\gamma} & B_{\gamma\gamma} & 0 & 0 & 0 \\
 0              &    0             & B_x & 0 & 0 \\
 0              &    0             & 0 & B_y & 0 \\
 0              &    0             & 0 &  0 & B_z \\
\end{array}\right) \ ,
\end{equation}
and the gradient
\begin{equation}
\boldsymbol\nabla=\left(
{\textstyle \partial_\beta,
\frac{1}{\beta}\partial_\gamma,
\frac{\mathrm{i}\hat{L}^\prime_x}{2\beta\sin(\gamma-\frac{2\pi}{3})},
\frac{\mathrm{i}\hat{L}^\prime_y}{2\beta\sin(\gamma-\frac{4\pi}{3})},
\frac{\mathrm{i}\hat{L}^\prime_z}{2\beta\sin(\gamma)}
}\right)^t \ .
\end{equation}
\end{subequations}
It is important to note that the components of the collective
mass tensor as well as the potential depend on the deformation coordinates
$(\beta,\gamma)$ but not on the orientation of the intrinsic
system $(\boldsymbol\vartheta)$. The actual orientation of the system
is incorporated  by the angular momentum operators of the
intrinsic system $\hat{L}'_k=\hat{L}'_k(\boldsymbol\vartheta)$ 
($k\in(x,y,z)\equiv(1,2,3)$). 
The choice of the intrinsic system is not unique, which results in 
additional symmetry conditions:
\begin{subequations}
\begin{eqnarray}
(\beta,\gamma)   &=& \textstyle X(\beta,\gamma-k\frac{2\pi}{3}) \ ,
\\
B_k(\beta,\gamma) &=& \textstyle B_z(\beta,\gamma-k\frac{2\pi}{3}) 
\end{eqnarray}
\end{subequations}
with $X\in\left\{V,B_{\beta\beta},B_{\gamma\gamma},B_{\beta\gamma}\right\}$ which
have to be fulfilled necessarily by the collective parameter functions \cite{Kum67}.

\subsection{Expansion of Collective Potential and Masses}
Following similar lines to the rotationally invariant expansion of the
generalized Bohr-Hamiltonian in \cite{Kum67} a symmetry conforming, polynomial expansion
of the collective potential and mass tensor can be derived by appropriate 
coupling of the spherical tensors of coordinates ${\boldsymbol\alpha}$ and 
conjugate momenta $\boldsymbol\pi$.

To account for the fluctuations of the collective potential with deformation $\beta$
while assuming a smooth interpolation in the triaxial degree of freedom, the 
polynomial expansion of the collective potential $V$ is approximated up to
lowest order in $\gamma$ while maintaining symmetry requirements and the most 
general $\beta$-dependency similar to \cite{Fle04a}:
\begin{equation}\label{eqn:PotExp}
V(\beta,\gamma) = V_1(\beta) + \beta^3\cos(3\gamma)V_1(\beta) \ .
\end{equation}
By means of similar arguments the $\gamma$-smooth parameterization of the 
collective mass tensor is given in (\ref{eqn:MassExp}). 
To end up with a polynomial expansion the functions 
$V_i(\beta), B_i(\beta)$ have to be analytic and even functions. 

%
\begin{figure*}[ht]
\begin{eqnarray}
B_{\beta\beta}(\beta,\gamma)   &=&
                                          B_1(\beta)           
                    +\beta  \cos(3\gamma) B_2(\beta)  
                    +\beta^2              B_3(\beta) 
                    +\beta^3\cos(3\gamma) B_4(\beta) \ ,
\nonumber\\
B_{\beta\gamma}(\beta,\gamma)  &=&
                     \hspace{0.7cm} 
                    -\beta\sin(3\gamma) B_2(\beta) 
                     \hspace{1.0cm}        
                    -\frac{5\beta^3}{12}\sin(3\gamma) B_4(\beta) \ , 
\nonumber\\
B_{\gamma\gamma}(\beta,\gamma) &=&
                                                    B_1(\beta)             
                    -\beta\cos(3\gamma)             B_2(\beta)  
                    +\frac{\beta^2}{6}              B_3(\beta) 
                    +\frac{\beta^3}{6}\cos(3\gamma) B_4(\beta)  \ ,
\nonumber\\[2.0ex]
B_{z}(\beta,\gamma)            &=&
                                                      B_1(\beta)  
                    -       \beta       \cos( \gamma) B_2(\beta) 
                    - \frac{\beta^2}{18}\left(7-10\cos(2\gamma)\right) B_3(\beta) 
                    - \frac{\beta^3}{18}\left(7\cos(3\gamma)-10\cos(\gamma)\right) B_4(\beta) \ .
\label{eqn:MassExp}
\end{eqnarray}
\end{figure*}
%
%
%
%
%

\subsection{Interpolation Formulas}
The parameterization of the collective potential 
(\ref{eqn:PotExp}) and mass (\ref{eqn:MassExp}) are
adjusted to the values obtained by
CHF and cranking along the axially symmetric cuts 
($\alpha^0=\alpha_{20}, \alpha^{\pm 2}=\alpha^{\pm 1}=0$). Besides the
Born-Oppenheimer surface $V(\alpha_{20})$ the three dimensional 
axial GCM-GOA Hamiltonian is defined from the $\beta$-vibrational
mass $B_{\beta\beta}(\alpha_{20})$ corresponding to motion along the
axial symmetric path and the degenerate rotational mass
$B_{x/y}(\alpha_{20})$ related to rotations around the
 non-symmetry axes. Generalizing the collective Hamiltonian to the
five dimensional configuration space yields the simple interpolation 
rules between the expansion coefficients in (\ref{eqn:MassExp})
and the collective 
parameter functions obtained from the axial calculations:
\begin{subequations}
\begin{eqnarray}
V_0(\beta) &=& \frac{V^{\rm(p)} + V^{\rm(o)}}{2} \ ,
\\
V_1(\beta) &=& \frac{V^{\rm(p)} - V^{\rm(o)}}{2\beta^3} \ , 
\\[2.0ex]
B_1(\beta) &=& \frac{1}{10}\left[
      2B_{\beta\beta}^{\rm (p)}
    + 2B_{\beta\beta}^{\rm (o)}
    + 3B_{x/y}^{\rm (p)}
    + 3B_{x/y}^{\rm (o)} \right] ,
\\
B_2(\beta) &=& \frac{1}{7\beta}\left[
      2B_{\beta\beta}^{\rm (p)}
    - 2B_{\beta\beta}^{\rm (o)}
    + 3B_{x/y}^{\rm (p)}
    - 3B_{x/y}^{\rm (o)} \right] ,
\\
B_3(\beta) &=& \frac{3}{10\beta^2}\left[
      B_{\beta\beta}^{\rm (p)}
    + B_{\beta\beta}^{\rm (o)}
    - B_{x/y}^{\rm (p)}
    - B_{x/y}^{\rm (o)} \right] ,
\\
B_4(\beta) &=& \frac{3}{14\beta^3}\left[
      B_{\beta\beta}^{\rm (p)}
    - B_{\beta\beta}^{\rm (o)}
    - 2B_{x/y}^{\rm (p)}
    + 2B_{x/y}^{\rm (o)} \right] ,
\end{eqnarray} 
\end{subequations}
where $X^{\rm (p)} = X(\alpha_{20}=\beta)$ and 
$X^{\rm (o)} = X(\alpha_{20}=-\beta)$. By inserting 
the expansion functions
$B_i$ and $V_i$ into (\ref{eqn:MassExp})
the collective mass tensor and potential are found.

\subsection{Zero-Point Energies}
The zero-point energies (ZPE) are composed
from the interpolated potential
$V(\beta,\gamma)$, 
the mass tensor ${\bf B}(\beta,\gamma)$
and the GOA width tensor $\boldsymbol\lambda(\beta,\gamma)$. Thereby
the width tensor is interpolated in the same manner as the collective
mass tensor. In the-five dimensional case the zero-point energy correction
also includes terms from spurious ZPE in the additional degrees of freedom,
namely the $\gamma$-vibration and $z$-rotation mode as well as contributions
from the $\beta$-$\gamma$-coupling. The contributions from the kinetic ZPE are 
constructed straight forward from the parameter functions by
\begin{eqnarray}
E^{\rm ZPE}_{\rm kin} 
&=&\!
  \frac{1}{2}\lambda_{\mu\nu} B^{\mu\nu}
\\
&=&\! \frac{1}{2}
\!\!\left[
  \lambda_{\beta\beta}B_{\beta\beta} + 
  \lambda_{\gamma\gamma}B_{\gamma\gamma} +
2\lambda_{\beta\gamma}B_{\beta\gamma} +
  \!\!\!\!\!\sum_{k=x,y,z}\!\!\!\!\!\lambda_{k}B_{k} \right] \ .
\nonumber
\end{eqnarray}
A compact notation of the potential ZPE requires to define
the curvature tensor of the collective potential 
$C_{\mu\nu}=\frac{1}{2}\nabla_\mu\nabla_\nu V$ with intrinsic 
components
\begin{eqnarray}
C_{\beta\beta} &=& \frac{1}{2}
\partial^2_\beta V(\beta,\gamma) \ ,
\nonumber\\
C_{\gamma\gamma} &=&\frac{1}{2}
\left[\frac{1}{\beta}\partial_\beta + \frac{1}{\beta^2}\partial^2_\gamma\right] V(\beta,\gamma) \ ,
\nonumber\\
C_{\beta\gamma} &=&\frac{1}{2}
\left[\frac{1}{\beta}\partial_\beta\partial_\gamma - \frac{1}{\beta^2}\partial_\gamma\right] V(\beta,\gamma) \ ,
\nonumber\\
C_{k} &=&\frac{1}{2}
\left[\frac{1}{\beta}\partial_\beta + \frac{\cot(\gamma-k\frac{2\pi}{3})}{\beta^2}\partial_\gamma\right] V(\beta,\gamma) \ .
\end{eqnarray}
In terms of the components of the curvature tensor ${\bf C}$ the potential ZPE 
is evaluated in analogy to the kinetic ZPE and reads
\begin{eqnarray}
E^{\rm ZPE}_{\rm pot} 
&=&
  \!\frac{1}{2}(\lambda^{-1})^{\mu\nu} C_{\mu\nu}
\\
&=& \!\frac{1}{2}
\!\!\left[
  \lambda^{-1}_{\beta\beta}  C_{\beta\beta} + 
  \lambda^{-1}_{\gamma\gamma}C_{\gamma\gamma} +
 2\lambda^{-1}_{\beta\gamma} C_{\beta\gamma} +
  \!\!\!\!\!\sum_{k=x,y,z}\!\!\!\!\!\lambda^{-1}_{k}C_{k} \right] \ . 
\nonumber
\end{eqnarray}
The topological interpolation scheme
(\ref{eqn:PotExp}, \ref{eqn:MassExp}) enforces the
asymptotic behavior for the ZPEs when approaching the
spherical limit: As expected from the harmonic approximation
the contributions from the two
vibrational modes and the three rotational modes degenerate yielding
five equivalent contributions for $\beta=0$.
This feature of our model enables an improved 
description of ground state properties of magic and especially 
doubly-magic nuclei.

\end{appendix}
\bibliographystyle{epj}
\bibliography{biblio,reviews,books,add}

\end{document}